\documentclass[multphys,vecphys]{svmult}
\usepackage{makeidx}         
\usepackage{graphicx}        
\usepackage{multicol}        
\usepackage[bottom]{footmisc}
\makeindex             

\begin{document}
\title*{Astrophysical Probes of Fundamental Physics}
\author{C.J.A.P. Martins}
\institute{CFP, Universidade do Porto, Rua do Campo Alegre 687, 4169-007 Porto, Portugal\\
DAMTP, University of Cambridge, Wilberforce Rd., Cambridge CB3 0WA, U.K.
\texttt{C.J.A.P.Martins@damtp.cam.ac.uk}}
\maketitle

\begin{abstract}
I review the theoretical motivation for varying fundamental couplings and discuss how these measurements can be used to constrain a number of fundamental physics scenarios that would otherwise be inacessible to experiment. As a case study I will focus on the relation between varying couplings and dark energy, and explain how varying coupling measurements can be used to probe the nature of dark energy, with important advantages over the standard methods. Assuming that the current observational evidence for varying $\alpha$ and $\mu$ is correct, a several-sigma detection of dynamical dark energy is feasible within a few years, using currently operational ground-based facilities. With forthcoming instruments like CODEX, a high-accuracy reconstruction of the equation of state may be possible all the way up to redshift $z\sim4$.
\end{abstract}

\section{Theoretical Expectations}
\label{cjm:theory}

The deepest question of modern physics is whether or not there are fundamental scalar fields in nature. They are a key ingredient in the standard model of particle physics (cf. the Higgs particle, which is supposed to give mass to all other particles and make the theory gauge-invariant), but after four decades of particle physics searches there is still no evidence that nature has any use for them. Yet in recent years we have come to realize that the early universe is an ideal place to search for scalar fields, if they exist at all, and there have been some possible hints for them in various contexts. The field of astrophysical searches for varying couplings has bee particularly active in recent years, as can be seen by the extensive series of new results and ongoing or forthcoming projects presented at this conference. Observations suggest that the recent universe is dominated by an energy component whose gravitational behaviour is quite similar to that of a cosmological constant. This could of course be the right answer, but the observationally required value is so much smaller than what would be expected from particle physics that a dynamical scalar field is arguably a more likely explanation. Theoretical motivation for such a field is not hard to find. In string theory, for example, dimensionful parameters are expressed in terms of the string mass scale and a scalar field vacuum expectation value.

Now, the slow-roll of this field (which is mandatory so as to yield negative pressure) and the fact that it is presently dominating the universe imply (if the minimum of the potential vanishes) that the field vacuum expectation value today must be of order $m_{Pl}$, and that its excitations are very light, with $m\sim H_0\sim 10^{-33}$ eV. But a further consequence of this is seldom emphasized~\cite{cjm:CARROLL}: couplings of this field lead to observable long-range forces and to time-dependence of the constants of nature (with corresponding violations of the Einstein Equivalence Principle). A spacetime varying scalar field coupling to matter mediates a new interaction. If the recent evidence for varying couplings~\cite{cjm:MURPHY,cjm:UBACHS} is explained by a dynamical scalar field, this automatically implies the existence of a new force. A series of space missions (ACES, $\mu$SCOPE, STEP) will improve on current bounds on the Einstein Equivalence Principle by as many as 6 orders of magnitude. These must find violations if the current data is correct~\cite{cjm:DAMOUR}. Joint analyses of varying coupling and Equivalence Principle measurements will shortly provide key tests to a number of fundamental paradigms, such as string theory (and may well be our only opportunity to find evidence for it).

Moreover, in theories where a dynamical scalar field is responsible for varying $\alpha$, the other gauge and Yukawa couplings are also expected to vary. Specifically, in GUTs there is a relation between the variation of $\alpha$ and that of the QCD scale, $\Lambda_{QCD}$, implying that the nucleon mass will vary when measured in units of the Planck mass. Similarly, one would expect variations in the Higgs vacuum expectation value (VEV), $v$, leading to changes in all particle mass scales including the electron mass. We therefore expect variations of the proton-to-electron mass ratio, $\mu=m_p/m_e$. Measurements of $\mu$ have been suggested a long time ago~\cite{cjm:THOMPSON}. Typically one has~\cite{cjm:CALMET}
\begin{equation}
\frac{\dot\mu}{\mu}\sim\frac{\dot\Lambda_{QCD}}{\Lambda_{QCD}} - \frac{\dot{v}}{v} 
\sim R\frac{\dot \alpha}{\alpha}\,;\label{cjm:defineR}
\end{equation}
the latter equality should be seen as the first term in a Taylor series, but given the expected level of variations the approximation should be good enough for most purposes. The value of $R$ is model-dependent (indeed, even its sign is not determined \textit{a priori}), but large values and negative values are naively expected for GUT models in which modifications come from high-energy scales.  The large proportionality factors arise simply because the strong coupling constant and the Higgs VEV run (exponentially) faster than $\alpha$. Note that with current data \cite{cjm:MURPHY,cjm:UBACHS,cjm:Kanekar} one infers $R\sim-4$. Be that as it may, the wide range of $\alpha$-$\mu$ relations implies that simultaneous measurements of both are a powerful discriminating tool between competing models: we can in principle test GUT scenarios without ever needing to detect any GUT model particles, say at accelerators.

\section{From $\alpha$ and $\mu$ to $w(z)$}
\label{cjm:eos}

A crucial goal of modern observational cosmology is characterizing the properties of dark energy, and in particular to look for dynamical behaviour. A simple property is its equation of state, $w=p/\rho$, and considerable effort is being put into trying to measure it. Current methods of choice are type Ia supernovae and (more recently) weak lensing. However, the question arises as to whether these are indeed the best tools for the task at hand. It has been known for some time \cite{cjm:Maor1} that supernova measurements are limited as a probe of the dark energy equation of state, especially if it is varying with redshift. Analysis of current and future constraints on the dark energy equation of state, from the various standard approaches and parametrized in the usual way~\cite{cjm:Upadhye}, shows that a convincing detection of time variation of $w$ is quite unlikely even with hypothetical future space-based experiments such as DUNE or JDEM (in any of its many versions). This is expected since any dynamical field providing the dark energy must be slow-rolling at the present time, and for slow variations there will always be a constant $w$ model that produces nearly identical results over the redshift range where dark energy is dynamically important. This point has also recently been made in \cite{cjm:Liddle}, and even the Dark Energy Task Force report \cite{cjm:DETF} (an otherwise very naive document) revelas the shortcomings of the standard approaches.

\begin{figure}[t]
\centering
\includegraphics[height=7cm]{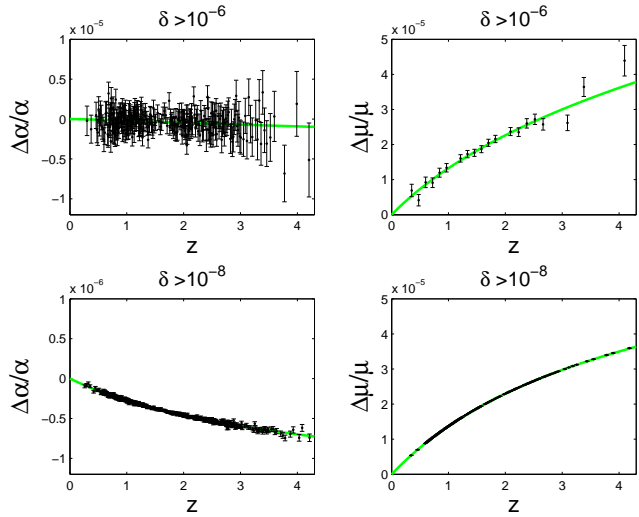}
\caption{Simulated datasets expected for $\alpha$ and $\mu$ in the near future (top panels) and with CODEX \protect\cite{cjm:Codex,cjm:Molaro} (lower panels), assuming a particular dark energy model.}
\label{cjm:fig1}
\end{figure}

Luckily, better (and cheaper) alternatives are available. A potentially effective tool for probing dynamical dark energy has been suggested previously in \cite{cjm:PARKINSON,cjm:NUNES}, though not yet studied in detail: probing varying couplings is a key test to these models, and in particular the varying couplings can be used to infer the evolution of the scalar field, and thus to determine its equation of state. This is analogous to reconstructing the 1D potential for the classical motion of a particle once its trajectory has been specified. Note that this reconstruction method requires only the calculation of first derivatives of noisy data, while the standard methods rely on claculating second derivatives of noisy data. Previous efforts only considered the variation of $\alpha$, but variations of $\mu$ may be easier to detect than those of $\alpha$ (if $R$ is indeed large), and thus provide tighter constraints on dark energy, although the number of such measurements is currently much smaller than the $\alpha$ dataset. (The main reason for this is the difficulty in finding molecular Hydrogen clouds.) One of the goals of our work \cite{cjm:OURMU} is to encourage further measurements of $\mu$, and significant such efforts are already in progress.

\begin{figure}[t]
\centering
\includegraphics[height=7cm]{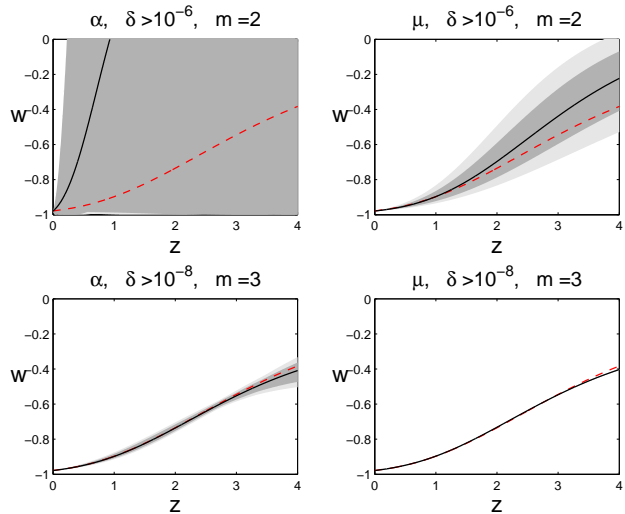}
\caption{The reconstruction of the equation of state and its error band is shown for the datasets of Fig. \protect\ref{cjm:fig1}. The dashed line represents the dark energy equation of state corresponding to the potential used to generate the simulated data and the solid line corresponds to the reconstruction's best fit. The dark and light region are the $1\sigma$ and $2\sigma$ confidence levels.}
\label{cjm:fig2}
\end{figure}

Having impreved measurements of both $\alpha$ and $\mu$ is extremely useful for various reasons. With both observables, the reconstruction will be a lot easier, not to mention less model-dependent. One has the advantage of a much larger lever arm in terms of redshift, since such measurements can be made up to redshifts of $z\sim4$. Naively one might think that this is not a big advantage, since dark energy is only dynamically relevant at relatively low redshift, and even the DETF report \cite{cjm:DETF} explicitly claims that there's no advantage in probing high redshifts. However, this completely misses the point that the additional redshift coverage probes the otherwise unaccessible $z$ range where scalar field dynamics is expected to be fastest, deep in the matter era. This not only make the detection of any possible dynamics easier, but also reduces (and possibly elliminates) the model-dependence that is unavoidable in the standard methods (where parametrisations like $w=w_0+w_a (1-a)$ are dangerously naive). Last but not least, this method provides direct evidence distinguishing dynamical dark energy from a cosmological constant, which given the current data may be very challenging for the standard cosmological tests. Figs. \ref{cjm:fig1} and \ref{cjm:fig2} show an example of our recent work~\cite{cjm:OURMU} displaying the benefits of a reconstruction using data on both couplings.

Finlly, let us point out the fact that, because the reconstruction method requires calculating (first) derivatives of data, it is important to have a good redshift coverage. Therefore it is also important to have a method of measurement that can be applied to a large range of redshifts without changing systematics. Such a method does exist for measuring $\alpha$: it is the Sunyaev-Zel'dovich effect~\cite{cjm:OURALPHA}, whose redshift-independence makes it ideal. Current data already provides interesting bounds, and an improvement of several orders of magnitude is expected in the coming years, when throusands of clusters will be at hand for this task, for example from the Planck Surveyor.

\section{Conclusions}
\label{cjm:end}

The prospects for further, more accurate measurements of fundamental constants are definitely bright, as has been highlighted at this conference. The methods described above and other completely new ones that may be devised thus offer the real prospect of an accurate mapping of the cosmological evolution of the fine-structure constant, $\alpha(z)$, and the proton to electron mass ratio, $\mu(z)$. This may well prove to be the most exciting area of research in the coming years. The worse that can happen to cosmology is the scenario where a number of cosmological parameters are fixed by WMAP, then nothing new happens until Planck comes along and merely adds one digit to the precision of each already-known parameter. After that cosmology may well be dead: there will be little incentive to pushing research further to figure out what the next digit is. However, if in the meantime violations of the Equivalence Principle and/or varying fundamental constants are unambiguously confirmed, then one will (finally) have evidence for the existence of new physics---most likely in the form of scalar fields---in nature (which one may legitimately hope that Planck is able to probe) and an entirely new era begins.

The possibility of using varying couplings to reconstruct the equation of state of dark energy is particularly promising, especially if one obtains further measurements of $\mu$. Let us emphasize that this method can applied now, using existing ground-based facilities. Assuming that the current observational evidence for varying couplings (as discussed during this conference) is correct, a several-sigma detection of dynamical dark energy could be obtained with only a few hundred hours of observation on a VLT-class telescope---an extremely modest investment given the potential gains. Let us stress once again the crucial advantage of a much larger lever arm in terms of redshift, since such measurements can easily be made up to redshifts of $z\sim 4$: this is perhaps the only way one can probe the redshift range where the field evolution is expected to be fastest (if it is a tracking field)---that is, deep in the matter era.

Last but not least, this is also an example of how astrophysical observations can be optimal probes of fundamental physics. Such astrophysical probes will become increasingly common in years to come, and hope this provides early encouragement for the observational astrophysics community. The early universe is the best possible laboratory for fundamental physics, and in this era of precision astrophysics and cosmology, astrophysics has the observational tools that can provide a unique impulse to the fundamental physics of this century. The opportunity is there, and one should take it.

\section*{Acknowledgments}
I am grateful to Pedro Avelino, Nelson Nunes ans Keith Olive for an enjoyable collaboration that led to \cite{cjm:OURMU}. I also acknowledge stimulating conversations on the topic of this article with Nissim Kanekar and Rodger Thompson. This work was funded in part by FCT (Portugal), in the framework of the POCI2010 program, supported by FEDER. Specific funding came from grant POCI/CTE-AST/60808/2004.

\printindex
\end{document}